\documentclass
[secnumarabic,amsmath,amssymb,showpacs,nobibnotes,aps, prl,12pt]{revtex4-1}

\usepackage{amssymb}

\usepackage{amssymb}

\usepackage{graphicx}
\usepackage{dcolumn}
\usepackage{bm}
\usepackage{color}

\begin{document}
\title{On the kinetic theory of turbulence in plasmas}

\author{Shaojie Wang}
\email{wangsj@ustc.edu.cn}
\affiliation{Department of Modern Physics, University of Science and Technology of China, Hefei, 230026, China}
\date{\today}

\begin{abstract}
From the nonlinear (NL) Vlasov equation, a NL turbulence scattering term is found to describe the stochastic dissipation on the time scale longer than the turbulence correlation time. The evolution of the plasma distribution is determined by the well-understood unperturbed motion of charged particles, with the effects of fluctuating part of fields described by the turbulence scattering term. In the new framework, one can identify various important physics, covering from the linear and quasilinear to the NL regime; in particular, the connections between the widely used Kadomtsev-Pogutse's equation and Frieman-Chen's equation. The NL scattering term indicates the Onsager symmetry relation of turbulent transport and a NL frequency or $\bm k$ spectrum shift of a resonantly excited wave.
\end{abstract}

\pacs{52.25.Dg, 52.25.Fi, 52.20.Dq, 52.65.-y}

\maketitle

I. INTRODUCTION

Plasma transport is one of the most challenging scientific problems. It is theoretically treated by solving the kinetic equation that determines the evolution of the particle distribution in the phase space, and the kinetic equation is nonlinearly coupled with Maxwell's equations, because the motion of charged particles depends on the electric magnetic fields and meanwhile the fields depend on the distribution of the charged particles \cite{HazeltineBook98, HintonRMP76, HintonBook83, HirshmanNF81,  KadomtsevBook70, BalescuBook05, DiamondBook10}. Generally,  the fields can be decomposed into the fluctuating part and the averaged part \cite{HazeltineBook98}, and the motion of charged particles in averaged fields is usually well-understood. The fluctuation of fields can be due to the particle discreteness and the collective instabilities of the plasma, which are known as collisional process \cite{HintonRMP76, HintonBook83, HirshmanNF81} and micro-turbulence \cite{ KadomtsevBook70, BalescuBook05, DiamondBook10}, respectively. Macroscopically, both the collisional transport \cite{HintonRMP76, HintonBook83, HirshmanNF81} and the turbulent transport \cite{EscandePRL07, WangPoP12, BalescuBook05, DiamondBook10} can be described by diffusion-convection fluxes.

Up to the present, the collisional transport and the turbulent transport are theoretically investigated in different ways. In the well-established collisional transport theory \cite{HintonRMP76, HirshmanNF81}, the particle motion due to the fields fluctuation induced by the particle discreteness is decoupled from the unperturbed motion in the averaged fields \cite{HazeltineBook98}, and the decoupled motion is described by the Landau collision operator \cite{HintonBook83}, which explicitly reveals the dissipation on the time scale longer than the typical correlation time in the binary collisional process. In the turbulent transport theory and numerical simulation, however, the particle motion due to the field fluctuation is usually not decoupled, and the particle trajectory has to be evaluated with the full fields \cite {BalescuBook05,DiamondBook10}. In this way, one has to solve the nonlinear (NL) Vlasov equation or its gyro-averaged version for the magnetic confinement fusion plasma, by following the particle trajectory in the full fields \cite{FriemanPF82, HahmPF88b, BrizardRMP07, BrizardJPP89,LinScience98, JenkoPoP00, CandyJCP03, SatoCSD09, GarbetNF10}.
In contrast to the usual NL gyro-averaged Vlasov equation \cite{FriemanPF82} without the turbulence dissipation term explicitly revealed, Kadomtsev and Pogutse \cite{KadomtsevBook70} heuristically introduced a turbulent thermal conduction term into the drift kinetic equation to discuss the NL turbulent transport (dissipation), which is also a traditional challenging problem in turbulence of neutral fluids \cite{FrischBook95}.

Recently, it has been proposed that the particle motion in the fluctuation fields due to the collective process can be decoupled from the unperturbed motion, by using the Hamiltonian Lie-transform perturbation method, and the quasilinear (QL) transport theory has been successfully demonstrated \cite{WangPoP12}. In this paper, by using the new method, we will show that, in a NL turbulent plasma, the particle motion due to the fluctuating part of fields can be decoupled from the well-understood unperturbed motion, and the effects of fluctuating fields on the particle distribution can be put into a turbulence scattering (TS) term, which explicitly reveals the NL stochastic dissipation on the time scale longer than the turbulence correlation time. The turbulent and the collisional transport can be theoretically treated in a similar way. The turbulence (NL) scattering term provides a new framework in first principle for understanding the physical kinetics of turbulence and NL behavior of coherent modes, which are hidden in the NL Vlasov equation widely used in physics community including fusion plasma instability and turbulence \cite{KadomtsevBook70, FriemanPF82}, Alfven turbulence in space and astrophysical plasmas \cite{ChenJGR99}, and NL optics \cite{HallPRE02,SilvaOC01}.

We begin with the general kinetic theory of plasma transport \cite{HazeltineBook98}.
\begin{subequations}
\begin{eqnarray}
&&\left[\partial_t+\bm v\cdot\partial_{ \bm x}+\frac{e_s}{m_s}\left(\bm E+ \bm v\times\bm B\right)\cdot\partial_{ \bm v}\right]f_s=0. \label{eq:vp}\\
&&\nabla\cdot\bm E=\frac{1}{\epsilon_0}\left(\rho_{ex}+\Sigma_s\int d^3 \bm v e_sf_s\right), \label{eq:poisson}\\
&&\nabla\times\bm B=\mu_0\left(\bm j_{ex}+\Sigma_s \int d^3 \bm v e_s \bm v f_s\right)+\frac{1}{c^2}\partial_t \bm E, \label{eq:ampere}\\
&&\nabla\times \bm E=-\partial_t \bm B,\\
&&\nabla\cdot\bm B=0,
\end{eqnarray}
\end{subequations}
where $f_s(\bm x, \bm v, t)$ is the distribution function of particle species $s$ at time $t$, with $\bm x$ and $\bm v$ the particle position and velocity, respectively; $e_s$ and $m_s$ the charge and mass of the particle, respectively; $\bm E$ and $\bm B$ the electric and magnetic fields, respectively. $\rho_{ex}$ and $\bm j_{ex}$ are electrical charge density and current density due to the external sources, respectively. In the following, the subscript $s$ shall be omitted for simplicity.

For comparison, we write down the kinetic equation for collisional transport \cite{HintonBook83}.
\begin{subequations}
\begin{eqnarray}
 \mathcal V_0 \left(f\right)&=&\mathcal C \left(f\right)\equiv
 -\partial_{ \bm v}\cdot\left[\left(\bm a_{\mathcal C}-\frac{1}{2}\bm d_{\mathcal C}\cdot\partial_{ \bm v}\right)f\right],\label{eq:C}\\
 \mathcal V_0 &\equiv& \partial_t+\bm{v}\cdot\partial_{ \bm x}+\frac{e}{m}\left(\bm E_0+ \bm v\times\bm B_0\right)\cdot\partial_{ \bm v},
\end{eqnarray}
\end{subequations}
where the velocity space convection vector $\bm a_{\mathcal C}$ and diffusion tensor $\bm d_{\mathcal C}$ depend on \cite{HintonBook83} $f$. $\bm E_0$ and $\bm B_0$ are averaged fields, with the fluctuation due to particle discreteness removed to the collision term, $\mathcal C \left(f\right)$. Usually, the effects of the micro-instabilities are not included in the collisional transport theory.

II. I-TRANSFORM

In the following, to concentrate on the turbulent transport, we shall deal with Eq. (\ref{eq:vp}), with $\bm E=\bm E_0+\delta \bm E$ and $\bm B=\bm B_0+\delta \bm B$ understood as the fields without fluctuation due to the particle discreteness, and $\bm E_0$ and $\bm B_0$ understood as the equilibrium fields that do not include the fluctuations $(\delta \bm E, \delta \bm B)$ due to micro-instabilities.

We begin with the particle motion in the unperturbed electromagnetic fields. The unperturbed fundamental one-form (phase space Lagrangian) \cite{CaryRMP09} is written in terms of the noncanonical variables $\bm z=(\bm x,\bm v)$ as
\begin{equation}
 \gamma_{0}= \left[m \bm v+e \bm A_0\left(\bm x\right)\right]\cdot d\bm x-H_0\left(\bm x,\bm v\right)dt,
\end{equation}
with the unperturbed Hamiltonian $H_0=(1/2)m|\bm v|^2+e\Phi_0(\bm x)$.  $\bm E_0=-\nabla\Phi_0$. $\bm B_0=\nabla\times \bm A_0$. The well-understood equations of unperturbed motion are
\begin{equation}
\dot z_0^i=\{z^i,H_0\}=J_0^{ij}\partial_j H_0\equiv \{z^i,z^j\}\partial_j H_0,\label {eq:Z0dot}
\end{equation}
with the unperturbed Poisson bracket \cite{CaryRMP09}
\begin{equation}
\{f,g\}=\frac{1}{m}\left(\partial_{\bm x} f\cdot \partial_{\bm v}g- \partial_{\bm v}f\cdot\partial_{\bm x} g\right)-\frac{e}{m^2}\bm B_0 \cdot\partial_{\bm v}f\times\partial_{\bm v}g.\label{eq:PB}
\end{equation}

Introduce the perturbations of the electric field $\delta \bm E=-\nabla \delta \Phi(\bm x,t)-\partial_t\delta \bm A (\bm x,t)$ and the magnetic field $\delta \bm B=\nabla\times \delta \bm A$, which are due to the micro-instabilities.  The fundamental one-form is separated into the unperturbed part and the perturbation part, $\gamma = \gamma_0+\gamma_1$, with
\begin{equation}
\gamma_1  = e\delta \bm A(\bm x,t)\cdot d\bm x -e\delta \Phi(\bm x,t)dt.
\end{equation}

We shall endeavor to decouple the perturbed motion due to the turbulence from the unperturbed motion by using the recently proposed method of I-transform \cite{WangPoP12}, which, as a special Lie-transform \cite{CaryAoP83, LittlejohnJMP82}, makes the equations of motion in the perturbed fields formally identical to the unperturbed motion. The ordering parameter is $\epsilon_\delta\sim |\delta \bm E / \bm B_0 v_{th}|\sim|\delta \bm B/\bm B_0|\ll1$, with $v_{th}$ the thermal speed of the particle. The strong turbulence case with $\epsilon_\delta\sim 1$ is beyond the scope of this paper.

To $\mathcal O(\epsilon_\delta^2)$, the phase-space Lie transform \cite{LittlejohnJMP82,CaryAoP83} between the old variables $\bm z$ and the new variables $\bar {\bm z}$ is
\begin{subequations}
\begin{eqnarray}
\bar z^i &=& z^i + g_1^i(\bm z) + g_2^i + \frac{1}{2}g_1^j\partial_j g_1^i, \label{eq:barZ}\\
 z^i &=& \bar z^i - g_1^i(\bar {\bm z}) - g_2^i + \frac{1}{2}g_1^j\partial_j g_1^i, \label{eq:Z}
\end{eqnarray}
\end{subequations}
where $\bm g_1$ and $\bm g_2$ are the first and second order Lie-transform generating vectors, respectively.

The distribution function (a scalar) is transformed according to
$\bar f \left(\bar {\bm z}\right)= f\left(\bm z\right)$; to $\mathcal O(\epsilon_\delta^2)$,
\begin{subequations}
\begin{eqnarray}
\bar f &=& f -\left(g_1^i +  g_2^i\right)\partial_i f + \frac{1}{2} g_1^i\partial_i  g_1^j\partial_j f, \label{eq:fpush}\\
f &=& \bar f +\left( g_1^i +  g_2^i\right)\partial_i \bar f + \frac{1}{2} g_1^i\partial_i  g_1^j\partial_j \bar f. \label{eq:fpull}
\end{eqnarray}
\end{subequations}

The straightforward calculation of the I-transform beginning with $\gamma_0$ and $\gamma_1$ is similar to Ref. \onlinecite{WangPoP12}. The results needed are summarized as follows.

The I-transform gauge functions, $s_n$, are solved by
\begin{equation}
\mathcal V_0\left(s_n\right)=e\left(\delta \Psi_n-\bm v\cdot \delta\bm{\mathcal A}_n\right).\label{eq:Sn}
\end{equation}

The I-transform generating vectors can be written as
\begin{subequations}
\begin{eqnarray}
g_n^{\bm x}&=&-\frac{1}{m}\partial_{\bm v} s_n,\label{eq:Gxn} \\
g_n^{\bm v}&=&\frac{1}{m}\partial_{\bm x} s_n+\frac{e}{m^2}\bm B_0\times\partial_{\bm v} s_n+\frac{e}{m}\delta\bm{\mathcal A}_n. \label{eq:Gvn}
\end{eqnarray}
\end{subequations}

The effective potentials used above are given by
\begin{subequations}
\begin{eqnarray}
&&[\delta \bm {\mathcal A}_1,\delta {\Psi}_1]=[\delta \bm A, \delta \Phi],\\
&&[\delta \bm {\mathcal A}_2,\delta {\Psi}_2]=
 [ \frac{1}{2} g_1^{\bm x} \times \delta \bm B, \frac{1}{2} g_1^{\bm x}\cdot \delta \bm E ].
\end{eqnarray}
\end{subequations}

With the generating vectors, the I-transformed fundamental one-form is identical to the unperturbed one,
\begin{equation}
\bar {\gamma_{0}}= \left[m \bar{\bm v}+e \bm A_0\left(\bar {\bm x}\right)\right]\cdot d \bar {\bm x}-H_0\left(\bar {\bm x},\bar{\bm v}\right)dt,
\end{equation}
and hence the equations of motion are simply
\begin{eqnarray}
\dot{\bar z}^i=\{\bar z^i,H_0\}=J_0^{ij}\partial_j H_0.\label {eq:Zibardot}
\end{eqnarray}
The Vlasov equation is transformed to
\begin{equation}
\mathcal V_0\left(\bar f\right)=0.\label{eq:Vfbar}
\end{equation}

It is not hard to show that $\bm g_n$'s are incompressible flows in the phase space \cite{WangPoP12},
\begin{equation}
\partial_{\bm z}\cdot \bm g_n=0.\label{eq:div}
\end{equation}

III. NONLINEAR TURBULENCE SCATTERING TERM

 Note that $\bm g_n$'s describe the excursion from the unperturbed orbit. In describing the TS, Eq. (\ref{eq:Sn}) is understood as the stochastic equations \cite{BalescuBook05}. Since the I-transform is a perturbation theory, its validity requires that the real orbit does not deviate much from the unperturbed one; this is clearly the case in the linear and QL theory \cite{WangPoP12}. However, in the NL stage, the real orbit may deviate far away from the unperturbed one in a long-time, especially for a resonant particle; this is the familiar secularity difficulty in the regular perturbation theory \cite {CaryAoP83}. To circumvent the secularity problem, one can apply the I-transform in a short time interval $\Delta t$, an infinitesimal I-transform; the long-time motion can be computed by applying the infinitesimal I-transform in a step-by-step way.

To evolve the particle distribution within a short time interval $\Delta t$, we denote the distribution function at $t-\Delta t$ as $f\left(\bm z,t-\Delta t\right)$. In applying the I-transform, one should keep in mind that it is the functions rather than the values that the Lie-transform operates on \cite{CaryAoP83, LittlejohnJMP82}.

Set $s_n\left(\bm z,t-\Delta t\right)=0$, we have $\bm g_n\left(\bm z, t-\Delta t\right)=0$. This means that we do not make any transform at $t-\Delta t$; therefore, according to Eqs. (\ref{eq:fpush}, \ref{eq:barZ}), we have
\begin{equation}
\bar f\left(\bar {\bm z}, t-\Delta t\right)=f\left(\bar {\bm z}, t-\Delta t\right)=f\left(\bm z, t-\Delta t\right). \label{eq:fbart-}
\end{equation}

Integrating Eq. (\ref{eq:Sn}) along the unperturbed orbit from $t-\Delta t$ to $t$, one finds $s_n\left(\bm z, t\right)$ and $\bm g_n\left(\bm z, t\right)$.
Integrating Eq. (\ref{eq:Vfbar}) from $t-\Delta t$ to $t$ and using Eq. (\ref{eq:fbart-}), we found
\begin{equation}
\bar f \left(\bar {\bm z}, t\right)= \bar f \left[ \bar {\bm z}-\Delta \bm z_0, t-\Delta t \right]=f\left[\bar {\bm z}-\Delta \bm z_0, t-\Delta t\right],\label{eq:fbart}
\end{equation}
where $\bar{\bm z}-\Delta \bm z_0\left(\bar{\bm z}\right)$ denotes the phase space point that arrives at $\bar{\bm z}$ through running along the unperturbed orbit from $t-\Delta t$ to $t$; note that $\Delta \bm z_0\sim \Delta t$.

Using Eqs. (\ref{eq:fpull}, \ref{eq:div}), we found from Eq. (\ref{eq:fbart})
\begin{equation}
f\left(\bm z, t\right)=\bar f \left[ \bm z-\Delta \bm z_0, t-\Delta t \right]+\partial_{\bm z}\cdot\left[\left(\bm g_1+\bm g_2+\frac{1}{2}\bm g_1\bm g_1\cdot \partial_{\bm z}\right)\bar f\right].\label{eq:ftpull}
\end{equation}
With only one time application of the I-transform, one may combine Eq. (\ref{eq:fbart-}) and Eq. (\ref{eq:ftpull}) to find
\begin{equation}
f\left(\bm z, t\right)= f \left[ \bm z-\Delta \bm z_0, t-\Delta t \right]+\partial_{\bm z}\cdot\left[\left(\bm g_1+\bm g_2+\frac{1}{2}\bm g_1\bm g_1\cdot \partial_{\bm z}\right) f\right].\label{eq:fq1}
\end{equation}

To evolve the particle distribution in a long time, we repeatedly apply the infinitesimal I-transform. In the next step from $t$ to $t+\Delta t$, we set again $s_n\left(\bm z, t\right)=0=\bm g_n\left(\bm z, t\right)$, and in the way similar to Eq. (\ref{eq:fbart-}), we have
\begin{equation}
\bar f\left(\bar {\bm z}, t\right)=f\left(\bar {\bm z}, t\right). \label{eq:fbart-t}
\end{equation}

Collecting Eq. (\ref{eq:fbart-t}) and Eq. (\ref{eq:ftpull}), we found
\begin{equation}
 \mathcal V_0 \left(\bar f\right)=\mathcal T \left(\bar f\right)
 \equiv
 -\partial_{ \bar {\bm z}}\cdot\left[\left(\bm a_{\mathcal T}-\frac{1}{2}\bm d_{\mathcal T}\cdot\partial_{ \bar {\bm z}}\right)\bar f\right], \label{eq:TS}
\end{equation}
the long-time evolution equation; the phase space convection vector $\bm a_{\mathcal T}$ and the symmetric diffusion tensor $\bm d_{\mathcal T}$ used in the NL TS term $\mathcal T $ are given by
\begin{subequations}
\begin{eqnarray}
 a_{\mathcal T}^{i}&=&a_{{\mathcal T},1}^{i}+a_{{\mathcal T},2}^{i}\equiv\frac{-g_1^i}{\Delta t}+\frac{-g_2^i}{\Delta t},\\
 d_{\mathcal T}^{ij}&=&\frac{g_1^i g_1^j}{\Delta t}.\label{eq:dT}
\end{eqnarray}
\end{subequations}

Finally $f(\bm z)$ is found by using Eqs. (\ref{eq:fpull}, \ref{eq:div}),
\begin{equation}
f\left(\bm z\right)=\bar f \left( \bm z\right)+\partial_{\bm z}\cdot\left[\left(\bm g_1+\bm g_2+\frac{1}{2}\bm g_1\bm g_1\cdot \partial_{\bm z}\right)\bar f\right].\label{eq:fp}
\end{equation}
Note that Eq. (\ref{eq:fq1}) can be recovered by Eqs. (\ref{eq:TS}, \ref{eq:fp}).

The finiteness of $\bm d_{\mathcal T}$ and $\bm a_{\mathcal T,2}$ in the TS term is due to the stochasticity developed by the nonlinearity on the time scale $\Delta t$; similarly, the finiteness of $\bm d_{\mathcal C}$ in the collision term is due to the randomness of the collisional scattering events on the time scale longer than the typical correlation time in collisional process \cite{HazeltineBook98}. At this point, the scale of $\Delta t$ should be further clarified in connection with the NL stochasticity \cite{BalescuBook05}. Let $\delta t$ be the turbulence correlation time. Note the fact that the stochasticity displays on a time scale longer than $\delta t$, while in a time scale shorter than $\delta t$, the stochasticity does not explicitly display itself. If one looks on a time scale $\Delta t >\delta t$, the randomness due to the stochasticity makes $\bm d_{\mathcal T}$ and $\bm a_{\mathcal T,2}$ finite ($\bm g_1\bm g_1\sim \Delta t$); if one looks on the time scale $\Delta t \ll\delta t$, $\bm d_{\mathcal T}$ and $\bm a_{\mathcal T,2}$ are indeed unimportant ($\bm g_1\bm g_1\sim \Delta t^2$). Clearly, $\bm d_{\mathcal T}$ and $\bm a_{\mathcal T,2}$ describe the turbulent dissipation on the time scale longer than $\delta t$. In fact, if one determines to resolve the time scale much shorter than $\delta t$, $\bm d_{\mathcal T}$ and $\bm a_{\mathcal T,2}$  can be ignored, and the $\bm a_{\mathcal T,1}\cdot\partial_{\bar{\bm z}}\bar f$ term retained in Eq. (\ref{eq:TS}) describes all the nonlinearity; this is essentially the underlying physics behind Frieman-Chen's equation \cite{FriemanPF82} in gyro-averaged form; note that on this time scale, $\bm a_{\mathcal T,1}$ is the first order velocity in phase space. The robustness of the TS term shall be demonstrated in the following applications. To do this we separate the distribution function into the fluctuation part $\tilde f$ and the ensemble averaged part $\langle f\rangle$, that is $f=\langle f \rangle+\tilde f$, $\langle \tilde f \rangle=0$. Note that $\langle \bm a_{{\mathcal T},1} \rangle=0$, and $\langle f \rangle$ evolves on the transport time scale.

IV. APPLICATIONS

For the linear and QL theory, we may apply the I-transform only one time, and the solution is given by Eq. (\ref{eq:fq1}); according to the familiar linear and QL theory \cite{DiamondBook10} of instability, we shall interpret $\Delta t$ as $\infty$ and invoke the causality condition to set $\tilde f\left(t\rightarrow -\infty\right)\rightarrow 0$; besides, $\langle f \rangle\left(t\rightarrow -\infty\right)$ is understood as a constant of unperturbed motion \cite{WangPoP12, WangPRE01}. The linear solution is readily found
\begin{equation}
\tilde f_L=\bm g_1\cdot\partial_{\bm z} \langle f\rangle.\label{linears}
\end{equation}
And the QL version is clearly given by
\begin{equation}
\partial_t \langle f\rangle+\partial_{\bm z} \cdot \left[- \frac{1}{2}\partial_t \langle \bm g_1 \bm g_1   \rangle \cdot \partial_{\bm z} \langle f\rangle \right]=0.\label{quasilinears}
\end{equation}
This exactly recovers the previous linear and QL theory \cite{WangPoP12}.

The long-time behavior in the NL regime is governed by Eqs. (\ref{eq:TS}, \ref{eq:fp}). Note that the second term in the right-hand side of Eq. (\ref{eq:fp}) should be retained in computing the moments of $f$ used in Maxwell's equations. However, since it is always a small correction term, it is unimportant in the following discussions on the long-time behavior; with this small term dropped, the over-bar symbols in Eq. (\ref{eq:TS}) can be simply ignored.
The NL transport equation is given by
\begin{eqnarray}
 \partial_t \langle f\rangle &+& \partial_{ \bm z}\cdot\left[ \left( \langle \bm a_{\mathcal T} \rangle -\frac{1}{2}\langle\bm d_{\mathcal T}\rangle \cdot\partial_{ \bm z }\right) \langle f \rangle\right]\nonumber\\
 =&-&\partial_{ \bm z}\cdot\left[ \langle \widetilde {\bm a_{\mathcal T}} \tilde f -\frac{1}{2}\widetilde {\bm d_{\mathcal T}}\cdot\partial_{\bm z}\tilde f\rangle \right], \label{eq:NL0}
\end{eqnarray}
where the right-hand side reduces to the QL transport in the QL regime according to Eqs. (\ref{linears}, \ref{quasilinears}). Clearly, the QL transport is due to the phase coherence of the particle excursion $\bm g_1$ and the distribution fluctuation $\tilde f$; it describes the resonant diffusion. The NL diffusion tensor $\langle\bm d_{\mathcal T}\rangle$ and convection vector $\langle \bm a_{\mathcal T} \rangle $ are attributed to the stochastic TS effects. The resonant diffusion dominates in the QL regime, and the stochastic transport is important in the NL regime.

The NL evolution of the linearly unstable mode can be written as
\begin{equation}
 \left[\mathcal V_0+\langle\bm a_{{\mathcal T},2}\rangle\cdot\partial_{\bm z}\right] \tilde f=-\partial_{ \bm z}\cdot\left[ \bm a_{{\mathcal T},1}\langle f\rangle-\frac{1}{2}\langle\bm d_{\mathcal T}\rangle \cdot\partial_{ \bm z }\tilde f \right]+.... \label{eq:NL1}
\end{equation}
Note that a turbulent thermal conduction term [similar to $\langle\bm d_{\mathcal T}\rangle$] was heuristically added to the drift kinetic equation [similar to Eq. (\ref{eq:NL1})] by Kadomtsev and Pogutse \cite{KadomtsevBook70} to solve for the saturation level of $\tilde f$, by invoking Richardson-Kolmogorov's energy cascade argument \cite{FrischBook95}.

For simplicity, we introduce the eikonal ansatz, $\tilde f=f_{\bm k}e^{-i(\omega t-\bm k \cdot \bm x)}$. When the turbulence saturated, Eq. (\ref{eq:NL1}) should balance. The real terms are from the right-hand side; the first term $\sim\gamma_{L,\bm k} \tilde f$ is the linear driving term, with $\gamma_{L,\bm k}$ the linear growth rate; the second term $\sim-\langle\bm d_{\mathcal T}^{\bm x \bm x}\rangle k^2 \tilde f$ is the NL damping \cite{WessonBooK97} or the NL resonance broadening \cite{HazeltineBook98} due to the turbulent diffusion. Balancing the real terms results to the well-known approximate formula \cite{WessonBooK97, HazeltineBook98, KadomtsevBook70} of turbulent diffusion coefficient
\begin{equation}
 |\langle \bm d_{\mathcal T}^{\bm x \bm x}\rangle |\sim \gamma_{L,\bm k}/k^2, \label{eq:td}
\end{equation}
which can also be found by invoking Prandtl's "mixing-length" argument \cite{WessonBooK97, HazeltineBook98}. The discussions presented here and following Eq. (\ref{eq:fp}) indicate that the new TS term connects the mysterious dissipation term in Kadomtsev-Pogutse's equation to Frieman-Chen's equation.

The linear and QL theory have been recovered and the NL turbulence diffusion term has been systematically derived in the new framework. Besides, by using the proposed NL scattering term, one can easily make two important conclusions.
(1) Balancing the imaginary terms of Eq. (\ref{eq:NL1}) results to the NL resonance condition
\begin{equation}
\omega-\bm k\cdot\bm v-\bm k\cdot \left[\langle\bm a_{{\mathcal T},2}^{\bm x}\rangle- \frac{1}{2} \partial_{\bm z}\cdot \langle \bm d_{\mathcal T}^{\bm z \bm x} \rangle\right] \sim 0. \label{eq:nrc}
\end{equation}
This indicates a NL phase shift effect, which can be related to important topics widely discussed in NL optics \cite{SilvaOC01} and plasma physics \cite{MoralesPRL72,ZhangPRL12}: the NL frequency shift of a wave resonantly excited with a fixed $\bm k$, and the NL $\bm k$ spectrum shift in propagation of a wave excited with a fixed $\omega$. This effect may be related to the ponderomotive effects \cite{MoralesPRL74,WangPoP12}.
(2) The symmetry of the phase space QL diffusion tensor has been used to demonstrate the Onsager symmetry relation of QL transport \cite{WangPoP12}, as is similar to the Onsager relation of collisional transport due to the symmetry of collisional velocity space diffusion tensor \cite{HirshmanNF81, HintonRMP76}. Noting the symmetry of the turbulent diffusion tensor in phase space, one expects that the Onsager relation still holds in the NL regime, an important point for experimental data analysis.

V. DISCUSSIONS AND CONCLUSIONS

It is not hard to extend this theory to the gyrokinetic version that decouples the fast gyro-motion for the low-frequency turbulence \cite{BrizardRMP07, WangPoP12}. Note that the new method involves only integrating along the unperturbed orbit, instead of integrating along the full orbit in the conventional method. This may bring about advantages by making use of our knowledge of the well-understood unperturbed orbit, especially for the toroidal axisymmetric tokamak fusion system \cite{WangPoP99, WangPoP06, WangPRE01, XuPoP11}. This suggests a new method in nonlinear kinetic simulation of turbulence, which will be reported in a later publication.

In conclusion, the NL TS term has been found by using the I-transform method to decouple the turbulent motion from the unperturbed motion. The evolution of the plasma distribution is governed by the ballistic motion of charged particles along the orbit determined by the averaged electromagnetic fields, with the effects of fluctuating part of fields described by the NL scattering term. In the new framework, various important physics scattered in the literature can be identified in a unified way. Particularly, the proposed NL scattering term reveals the connections between the well-known Kadomtsev-Pogutse's equation with a mysterious dissipation term and Frieman-Chen's equation without the explicit dissipation term. The NL scattering term indicates the Onsager symmetry relation of NL turbulent transport and a NL phase-shift effect related to the NL frequency or $\bm k$ spectrum shift of a resonantly excited wave.

\begin{acknowledgments}
This work was supported by the National Natural Science Foundation of China under Grant No. 11175178.
\end{acknowledgments}

\nocite{*}


%

\end{document}